\documentclass[10pt,prb,aps,showpacs,groupedaddress,superscriptaddress,twocolumn]{revtex4-2}
\usepackage{mlmodern}
\usepackage{preamble}
\usepackage{booktabs}
\usepackage{hyphenat}
\begin{document}
	\title{\textsc{Adaptive Pseudoboson Density-Matrix Renormalization Group for Dilute 2D Systems}}
	\author{F.~J.~Pauw \orcidlink{0009-0006-4188-8503}}
	\affiliation{\ascaddress}
	\affiliation{\mcqstaddress}
	\author{T.~K\"ohler \orcidlink{0000-0003-3224-028X} \altaffiliation{These authors contributed equally to this work.}}
	\affiliation{\hwuaddress}
	\author{U.~Schollw\"ock \orcidlink{0000-0002-2538-1802}}
	\affiliation{\ascaddress}
	\affiliation{\mcqstaddress}
	\author{S.~Paeckel\orcidlink{0000-0001-8107-069X}}
	\affiliation{\ascaddress}
	\affiliation{\mcqstaddress}
	\date{\today}
	\begin{abstract}
		Simulating strongly correlated systems in two dimensions is notoriously challenging due to rapid entanglement growth and frustration.
		Here, we introduce the \gls{a3p-dmrg} tailored to explore the ground states of dilute lattice models.
		The method compresses cluster Hilbert spaces by retaining only the most probable low-occupation Fock states, identified via probabilistic bounds and refined through a self-consistent mean-field basis optimization.
		We demonstrate that \gls{a3p-dmrg} is advantageous in low-filling and weak-coupling regimes for large system sizes where conventional \acrshort{dmrg} struggles.
		This establishes the method as a versatile tool for studying dilute quantum many-body systems relevant to ultra-cold atom quantum simulators, photonic lattices, Moir\'e materials and quantum chemistry.
	\end{abstract}
	\maketitle
	\glsresetall
	\section{Introduction}
	The large-scale exploration of strongly correlated quantum many-body systems in two spatial dimensions remains one of the fundamental challenges in computational physics.
	Such systems host a wealth of emergent phenomena ranging from unconventional superconductivity~\cite{Bednorz1986,Kastner1998,Scalapino2012} to topologically ordered states~\cite{Tsui1982,Laughlin1983,Stormer1999,Han2012,Wen2017}, yet their numerical description is notoriously difficult.
	Exact diagonalization is restricted to very small system sizes, while quantum Monte Carlo methods suffer from the fermionic sign problem or, in bosonic systems, from sign problems induced by frustration~\cite{Foulkes2001,Troyer2005}.
	Consequently, tensor network methods have become indispensable tools for tackling interacting lattice models in low dimensions.
	\gls{mps}, the foundation of the \gls{dmrg}, are extremely successful in one dimension, where they exploit limited entanglement growth dictated by the area law~\cite{White1992,Schollwock2005,Schollwock2011}. 
	For the same reason, their application to two-dimensional systems is severely restricted, since the entanglement entropy grows linearly with the system width, implying an exponential growth of the required bond dimension~\cite{Vidal2003,Latorre2004,Eisert2010}.
	While \gls{ttn}~\cite{Shi2006,Tagliacozzo2009,Murg2010,Nakatani2013,Gerster2014} and \gls{peps}~\cite{Verstraete2004b,Verstraete2008,Murg2007,Cirac_2021} naturally extend tensor network states to two dimensions, they each face intrinsic limitations:
	in \gls{ttn}, the tree geometry constrains the representable entanglement in a bipartition-dependent manner, while \gls{peps} are limited by their high algorithmic complexity and the absence of an exact contraction scheme.
	This motivates the search for alternative representations, which exploit additional physical structure.
	\begin{figure}[!ht]
		 \includegraphics{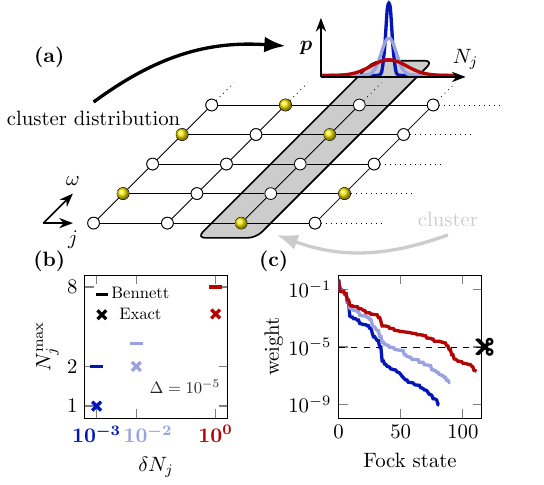}
		\caption{
			\textbf{Influence of cluster fluctuations on the effective many-body basis}
			\textbf{(a)} Square lattice of hard-core bosons (yellow spheres). 
			The gray box indicates a cluster of $W$ sites.
			The inset plot illustrates the probability distribution of the cluster-particle number for ground states with varying particle-fluctuations $\delta N_j$ and fixed average particle number $\bar{N_j}$. 
			\textbf{(b)} Highest relevant occupation number $N_j^{\mathrm{max}}$ obtained from the Bennett bound (bars) and from exact calculations (crosses) for a cluster of size $W=10$ as a function of $\delta N_j$ and for $\bar{N_j}=\nicefrac{1}{2}$. 
			The discarded weight is fixed at $\Delta = 10^{-5}$.
			\textbf{(c)} Decay of Fock-state weights for the three different ground states with varying values of $\delta N$.
		}
		\label{fig:apriori}
	\end{figure}
	A particularly challenging subset of two-dimensional systems arises in the dilute regime, where \gls{dmrg} often struggles to converge even local observables such as the density~\cite{Dolfi2012}.
	Probabilistic arguments, however, suggest that the low-energy Hilbert space can be effectively captured by constructing appropriate bases on dilute subsystems~\cite{Bennett1962}.
	In this picture, the most probable low-particle-number configurations dominate the subsystem Fock space when fluctuations are strongly suppressed, as illustrated in~\cref{fig:apriori}.
	This drastically reduces the effective dimensionality of the problem in a suitably chosen many-body basis and naturally improves convergence.
	Motivated by these considerations, we introduce the \gls{3p-dmrg} method that reformulates two-dimensional lattice models in terms of renormalized many-body degrees of freedom which we refer to as \textit{pseudobosons}.
	In this representation, the physically relevant pseudobosons -- those Fock-states within the exponentially narrow particle-number corridor -- form the optimal truncated basis for describing the low-energy sector.
	While the pseudoboson ansatz reduces the two-dimensional problem to a one-dimensional chain of many-body degrees of freedom, it introduces two major challenges: 
	first, the mapping from local sites to pseudobosons breaks potential $\mathrm{U}(1)$ particle-number symmetry, and second, it can lead to large effective local Hilbert-space dimensions.
	These drawbacks can be mitigated by combining pseudobosons with \gls{pp}~\cite{Stolpp2021,Koehler2021}, which both restores control over symmetries and allows to truncate the effective many-body Hilbert space systematically during a DMRG sweep.
	Since suppressed fluctuations are often accompanied by weak inter-cluster correlations, the pseudoboson basis can be further refined via a self-consistent optimization scheme where the most relevant many-body Fock states identified in this process then define an optimized basis tailored to the correlations of the full two-dimensional system.
	We refer to this extension as \gls{a3p-dmrg}.
	In short, our self-consistent basis construction allows to adapt flexibly to varying interaction strengths, density regimes, and cluster couplings while preserving computational efficiency, enabling simulations of large lattices ($\mathcal{L} \geq 20 \times 20$) even at low filling.
	Importantly, the pseudoboson ansatz is not restricted to particle-number degrees of freedom, but, in fact, generalizes to arbitrary $\mathrm{U}(1)$ quantum numbers.
	Recently Ref.~\cite{Mardazad2025} introduced a conceptually similar approach where the local many-body basis is not chosen from a mean-field criterion but explicitly constructed from the low-energy cluster eigenbasis.
	The scenario of dilute and weakly-coupled clusters is not purely of theoretical interest but arises in a wide range of physical contexts.
	Ultra-cold atoms in optical lattices, for instance, allow precise control over particle filling, enabling the realization of interacting models in two dimensions, even in the presence of artificial magnetic fields~\cite{Gross2017,Altman2021,Goldman2016,Aidelsburger2013}. 
	In this setting, the weak-coupling limit of $W$-legged flux ladders~\cite{Lehur2015,Buser2020,Palm2022} is particularly relevant for probing the robustness of topological order and has direct experimental significance~\cite{Tai2017,Impertro2025}.
	Similarly, comb-like geometries of dissipative hard-core bosons can host transverse quantum fluids and potentially exhibit a superfluid-to-supersolid transition~\cite{Kuklov2024,Pollet2018}, though it remains unclear whether these features persist in the dilute, weak-coupling limit where large system sizes are required.
	The pseudoboson ansatz may also be useful for studying low-filling regimes of inhomogeneous \(t\text{-}J\) planes coupled solely through spin–spin interactions, which are inaccessible to other methods such as \gls{mps} plus mean-field approaches~\cite{Bollmark2023,Koehler2025}.
	Beyond that, \gls{a3p-dmrg} might also be of use for quantum chemistry applications where the inhomogeneous entanglement structure of weakly-correlated molecules consisting of strongly-correlated sub-parts naturally favors the many-body decomposition~\cite{Larsson2022,Ma2024}.
	This article is structured as follows.
	In~\cref{sec:Method} we introduce the pseudoboson ansatz and the technical details of the method.
	We then show how the microscopic details of the underlying model relate to the pseudoboson ansatz in~\cref{sec:Model}.
	Building on this analysis, we extend the method with an adaptive basis optimization in~\cref{sec:MF} and show extensive benchmark results for hard-core bosons with flux in~\cref{sec:Benchmarks}.
	We close with concluding remarks and an outlook in~\cref{sec:Conclusion}.
	\section{3P-DMRG}\label{sec:Method}
	Consider a two-dimensional lattice of size $\mathcal{L}=L\times W$ consisting of $L$ clusters with $W$ sites each.
	A lattice site is then labeled by the composite index $(j,\omega)$ with $j=1,\dots,L$ and $\omega=1,\dots,W$. 
	For simplicity, we consider local hard-core bosonic degrees of freedom with on-site Hilbert space
	\begin{equation}
		\mathcal{H}_{j,\omega} = \mathrm{span}\{ \ket{0}, \ket{1}\},
	\end{equation}
	where $\ket{0}$ and $\ket{1}$ denote empty and occupied states, respectively. 
	We will later remark on how one can generalize the method to soft-core bosonic or even fermionic degrees of freedom. 
	Creation and annihilation operators in this basis, $\hat{a}_{j,\omega}^\dagger$ and $\hat{a}^{\nodagger}_{j,\omega}$, obey the hard-core algebra, i.e.,
	\begin{equation}\label{eq:ccr}
		\{\hat{a}^{\nodagger}_{j,\omega}, \hat{a}_{j,\omega}^\dagger\} = 1, \qquad 
		(\hat{a}^{\nodagger}_{j,\omega})^2 = (\hat{a}_{j,\omega}^\dagger)^2 = 0,
	\end{equation}
	while operators acting on different sites commute.
	The corresponding number operator is $\hat{n}_{j,\omega} = \hat{a}_{j,\omega}^\dagger \hat{a}^{\nodagger}_{j,\omega}$.
	Any pure state on the lattice expressed in the hard-core bosonic occupation-number (Fock) basis reads
	\begin{equation}
		\ket{\psi} = \sum_{\boldsymbol{\sigma}} c^{\boldsymbol{\sigma}} \ket{\boldsymbol{\sigma}}, \ \boldsymbol{\sigma}=\left(\sigma_{\scriptscriptstyle{1,1}},\dots , \sigma_{\scriptscriptstyle{L,W}} \right),
	\end{equation}
	where $\boldsymbol{\sigma}$ runs over all possible $2^{\mathcal{L}}$ configurations characterized by the coefficient tensor $c^{\boldsymbol{\sigma}}$.
	For the $j$th cluster, a Fock-basis state can be generated by the ladder operators as
	\begin{equation}
		\ket{\sigma_{j,1},...,\sigma_{j,W}} = \prod_{\omega=1}^W(\hat{a}_{j,\omega}^\dagger)^{\sigma_{j,\omega}}\ket{\varnothing}.
	\end{equation}
	These basis states are (degenerate) eigenstates of the cluster number operator
	\begin{align}
		\hat{N}_j=\sum_{\omega=1}^W \hat{n}_{j,\omega},
	\end{align}
	with eigenvalues $N_j\in\{0,1,...,W\}$.
	Accordingly, the cluster Hilbert space $\mathcal{H}_j$ decomposes into a direct sum of Fock subspaces $\mathcal{F}_{\scriptscriptstyle{N_j}}$ associated with occupation-number sectors $N_j$:
	\begin{align}
		\mathcal{H}_j=\bigotimes_{\omega=1}^{W} \mathcal{H}_{j,\omega}=\bigoplus_{N_j=0}^{W}\mathcal{F}_{\scriptscriptstyle{N_j}}
	\end{align}
	where $\dim \mathcal{F}_{\scriptscriptstyle{N_j}}=\binom{W}{N_j}$.
	\subsection{A priori bound on relevant Fock-basis states}
	\begin{figure*}
		 \includegraphics{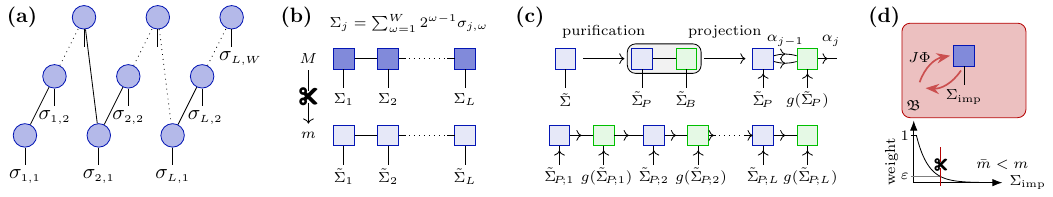}    
		\caption{\textbf{
				Schematic overview of the adaptive pseudobosonic \gls{mps} ansatz}
			\textbf{(a)} Snake-like mapping of the hard-core bosonic \gls{mps} onto the two-dimensional lattice.
			\textbf{(b)} A cluster of $W$ hard-core bosons is mapped onto a single pseudoboson with local dimension $M = 2^W$.
			The exponential suppression of relevant Fock-configurations enables truncation to an effective dimension $m$.
			\textbf{(c)} The broken $\mathrm{U}(1)$ symmetry is restored by projecting the purified pseudoboson states onto the physical submanifold satisfying $\tilde{\Sigma}_{P} + \tilde{\Sigma}_{B} = m$.
			\textbf{(d)} Improved pseudoboson basis of dimension $\bar{m}$ via self-consistent mean-field optimization scheme. 
		}
		\label{fig:overview}
	\end{figure*}
	\label{subsec:bennett_bound}
	By exploiting the Fock-space decomposition above, simple \textit{a priori} probabilistic arguments can be used to constrain the relevant size of the cluster Hilbert space $\mathcal{H}_j$.  
	To this end, we interpret the eigenvalue $N_j$ as a random variable, which itself is representable by the sum of the bounded and weakly correlated random variables $\sigma_{j,\omega}\in\{0,1\}$ under the probability measure induced by the reduced cluster density matrix $\hat{\rho}_j=\mathrm{Tr}_{k\neq j}\,|\Psi\rangle\langle\Psi|$ with respect to some target state $\ket{\Psi}$. 
	In this picture, \mbox{$p_{\scriptscriptstyle{N_j}}=\mathrm{Tr}(\hat{\rho}_j\,\hat{\mathcal{P}}_{\scriptscriptstyle{N_j}})$ }denotes the probability of finding $N_j$ particles on the cluster, where $\hat{\mathcal{P}}_{N_j}$ projects onto $\mathcal{F}_{\scriptscriptstyle{N_j}}$.
	Assume that both the expectation value \mbox{$\bar{N}_j=\sum_{\scriptscriptstyle{N_j}} N_j\,p_{\scriptscriptstyle{N_j}}$} and the fluctuation of the particle number \mbox{$\delta N_j =\sum_{\scriptscriptstyle{N_j}} (N_j-\bar{N}_j)^2 p_{\scriptscriptstyle{N_j}}$} are known.
	
	Then Bennett's inequality~\cite{Bennett1962} yields a one-sided upper bound on the probability of observing particle numbers exceeding the mean by at least $\mathcal{N}>0$
	\begin{equation}
		\underbrace{\mathbb{P} \big(N-\bar{N}\ge \mathcal{N}\big)}_{{\Delta}}
		\;\le\;
		\exp\!\left[-\,\delta N\,h\!\left(\frac{\mathcal{N}}{\delta N}\right)\right],
		\label{eq:bennett}
	\end{equation}
	where $h(u)=(1+u)\ln(1+u)-u$.
	Suppose that we fix a maximal discarded cumulative probability mass $\Delta\in(0,1)$. 
	By solving~\cref{eq:bennett} for $\mathcal{N}$, the set of particle numbers satisfying the bound is
	\begin{equation}
		\mathcal{I} = \{N\in\mathbb{N}_0 : N\leq N_j^{\mathrm{max}}\},
	\end{equation}
	where $N_j^{\mathrm{max}} = \lceil \mathcal{N} +\bar{N}\rceil$ is the smallest integer 
	greater than or equal to $\mathcal{N}+\bar{N}$ and constitutes the maximal relevant cluster-occupation number.
	Keeping only those Fock states with $N\in\mathcal{I}$ guarantees that the discarded probability mass -- and thus the total weight of all states outside these sectors -- does not exceed $\Delta$:
	\begin{equation}
		\sum_{N\notin \mathcal{I}} p_{\scriptscriptstyle{N}} = \mathbb{P}\big(N \geq N_j^{\mathrm{max}}\big) \;\le\; \Delta.
	\end{equation} 
	An \textit{a priori} upper bound on the effective local basis size thus follows as
	\begin{equation}
		\dim \mathcal{H}^{(\mathrm{kept})}
		\le
		\sum_{N\in \mathcal{I}} \dim\mathcal{F}_{\scriptscriptstyle{N}}
		=
		\sum_{N\in\mathcal{I}} \binom{W}{N},
		\label{eq:basis_bound}
	\end{equation}
	which depends only on the parameters $(\bar{N}, \delta N)$ and the tolerance $\Delta$ via~\cref{eq:bennett}, and scales polynomially in $W$ rather than exponentially.
	For example, achieving precision $\Delta\!\approx\!10^{-5}$ with moderate parameters \mbox{$(\bar{N}\!\approx\!\nicefrac{1}{2},\,\delta N\!=\!\nicefrac{1}{10})$} requires keeping Fock states up to $N\!=\!5$ occupations, independent of the cluster size.
	We confirm that the Bennett criterion indeed provides a strict upper bound for the effective Hilbert-space dimension for a cluster of $W=10$ sites for different values of $\delta N$, see~\cref{fig:apriori}(b).
	\subsection{Pseudoboson Representation}
	To exploit the probabilistic and physical constraints on the cluster particle sector, a suitable basis transformation from hard-core bosons to many-body Fock-basis states is required.
	Once again, consider a single cluster Hilbert space $\mathcal{H}_j$ of dimension 
	$\dim \mathcal{H}_j = 2^W \equiv M$ with Fock basis 
	$\{\ket{\sigma_{\scriptscriptstyle{j,1}},\dots,\sigma_{\scriptscriptstyle{j,W}}}\}$.
	By interpreting each Fock-basis state as a binary string and associating that string 
	with its corresponding integer representation (e.g., $\ket{100100}\leftrightarrow\ket{9}$), we can define a basis of $\mathcal{H}_j$ in terms of a single purely local effective site:
	\begin{align}
		\Bigl\{\ket{\Sigma_j},\ \Sigma_j\in\mathcal{M}=\{0,1,\dots,M-1\}\Bigr\}.
	\end{align}
	We refer to this composite degree of freedom as a \textit{pseudoboson}.
	The idea of large composite degrees of freedom is inspired by the \textit{pseudo-site} method introduced by Jeckelmann and White~\cite{Jeckelmann1998}.
	The one-to-one mapping between the $W=\log_2 M$ hard-core bosons and the pseudoboson with maximal occupation number $M-1$ is formally given by
	\begin{align} \label{eq:pbmapping}
		\Sigma_j&=\sum_{\omega=1}^{\log_2M}2^{\omega-1}\sigma_{\scriptscriptstyle{j,\omega}}, \\
		\sigma_{\scriptscriptstyle{j,\omega}}&=\lfloor\frac{\Sigma_j}{2^{\omega}}\rfloor \mod 2.
	\end{align}
	\begin{table}[b]%
		\caption{%
			Basis label transformation for a cluster of $W=3$ hard-core bosonic sites.
			$\tilde{\Sigma}$ denotes the effective pseudobosonic degree of freedom after a truncation of configurations with more than one occupied cluster site.
		}%
		\begin{tabular}{rccccccccc}%
			\toprule%
			& \multicolumn{8}{c}{state} & dimension \\ \midrule%
			$\sigma_{\scriptscriptstyle{\omega}}$: & 000 & 100 & 010 & 110 & 001 & 101 & 011 & 111 & $\mathcal{M}=8$ \\%
			$\Sigma\ $: & 0 & 1 & 2 & 3 & 4 & 5 & 6 & 7 & $\mathcal{M}=8$ \\%
			$\tilde{\Sigma}\ $: & 0 & 1 & 2 & - & 3 & - & - & - & $m=4$ \\%
			\bottomrule%
		\end{tabular}%
		\label{tab:mapping}%
	\end{table}%
	Assuming homogeneity, we will omit the cluster index $j$ from now on and work in the pseudoboson representation, but in principle the following calculations are readily generalizable to more complex systems with a less homogeneous structure.
	In ~\cref{tab:mapping} the mapping of the basis labels is summarized exemplarily for a cluster of size $W=3$. 
	Next, we express the fundamental operators in the pseudoboson basis. 
	The matrix elements of the hard-core bosonic ladder operators in the new basis are given by
	\begin{align}
		\hat{a}^{(\dagger)}_{\omega} = \sum_{\Sigma \in \mathcal{B}_{\omega}}\left( \ket{\Sigma - 2^{\omega-1}}\bra{\Sigma} \right)^{(\dagger)},
	\end{align}
	where $\mathcal{B}_{\omega} := \{ \Sigma \in \mathcal{M} \vert \sigma_{\omega}(\Sigma)=1 \}\subset \mathcal{M}$ is the set of all pseudobosonic labels corresponding to states 
	with a particle on site $\omega$ (i.e., $\sigma_{\omega}=1$).
	The corresponding occupation-number operator reads
	\begin{align}
		\hat{n}_{\omega} = \hat{a}_{\omega}^{\dagger}\hat{a}_{\omega}^{\nodagger}=\sum_{\Sigma \in \mathcal{B}_{\omega}}\ket{\Sigma}\bra{\Sigma}.
	\end{align}
	It is straightforward to verify that this representation satisfies the hard-core bosonic (anti)-commutation algebra (see~\cref{App:A}).
	Within the pseudoboson representation, the exponentially suppressed significance of many-body Fock states can be exploited to truncate the basis to an effective pseudobosonic degree of freedom $\tilde{\Sigma}$ with dimension $m=\sum_{N\in\mathcal{I}} \binom{W}{N}$, where $\mathcal{I}$ is determined by the Bennett bound.
	For the minimal example of a cluster with $W=3$ and a truncation of states with occupation number greater than one, the basis dimension is $m=W+1$, the set of labels for $\omega=1 $ is $\mathcal{B}_{1} =\{ 1 \}$, and the fundamental local operator reduces to $\hat{a}^{(\dagger)}_{1}=\left(\ket{0}\bra{1}\right)^{(\dagger)}$.
	The pseudoboson mapping in \cref{eq:pbmapping} can be readily extended to soft-core bosons and even fermions, see~\cref{App:B2}.
	For simplicity, in the remainder of this article, we will focus on the hard-core bosonic case.
	\subsection{Pseudoboson MPS}
	To utilize the pseudoboson structure in a \gls{dmrg} ground-state search, we formulate the corresponding pseudobosonic \gls{mps} representation.
	An arbitrary pure state on the two-dimensional lattice $\mathcal{L}$, expressed in the physical hard-core boson basis, takes the following \gls{mps} form
	\begin{align}
		\label{eq:mps}
		\ket{\psi}=\sum_{\sigma_1,...,\sigma_{\mathcal{L}}}A^{\sigma_1}_{\chi_0,\chi_1}\cdots A^{\sigma_{\mathcal{L}}}_{\chi_{\scriptscriptstyle{\mathcal{L}-1}},\chi_{\scriptscriptstyle{\mathcal{L}}}}\ket{\sigma_1,...,\sigma_{\mathcal{L}}},
	\end{align}
	where the tensor elements $A^{\sigma_j}\in\mathbb{C}^{\chi_{j-1}\times \chi_j}$ contain the variational parameters of the \gls{dmrg} algorithm. 
	The snake-like mapping of the two-dimensional lattice onto the intrinsically one-dimensional \gls{mps} is illustrated in~\cref{fig:overview}(a).
	Note that the MPS path is chosen such that the ordering of the tensors exploits the anisotropic entanglement distribution in the weak-cluster coupling limit.
	The complexity of \gls{dmrg} optimization of the MPS scales as $\sim 2 L \chi^3$, where  $\chi = \max_{j}\chi_j$ is the maximal bond dimension.
	The mapping from the hard-core bosonic \gls{mps} building blocks to the pseudobosonic ones is formally achieved by fusing the physical degrees of freedom $\{\sigma_{j,1},...,\sigma_{j,W}\}\rightarrow \{\Sigma_j\}$.
	To this end, $W$ rank-3 tensors $A$ are reshaped into a single dense rank-3 pseudoboson tensor $\mathbb{A}$, yielding
	\begin{align}
		A^{\sigma_{j,1}}\cdots A^{\sigma_{j,W}}=\mathbb{A}^{\Sigma_j}.
	\end{align}
	Consequently, in the pseudoboson basis the \gls{mps} representation of $\ket{\psi}$ reads
	\begin{align}
		\ket{\psi}=\sum_{\Sigma_1,..,\Sigma_{L}}\mathbb{A}_{\chi_0,\chi_1}^{\Sigma_1}\cdots \mathbb{A}_{\chi_{L-1},\chi_{L}}^{\Sigma_{L}}\ket{\Sigma_1,...,\Sigma_{L}},
	\end{align}
	(see~\cref{fig:overview}(b)) where in a \gls{dmrg} sweep we are optimizing over $\mathbb{A}_{\chi_{j-1},\chi_j}^{\Sigma_j}$.
	Importantly, the bond dimension $\chi$ connecting two adjacent clusters is preserved whereas the entanglement between sites within a cluster is naturally incorporated in the many-body formulation of the pseudobosons.
	Thus, treating a cluster of highly entangled sites -- which are only weakly entangled with neighboring clusters -- as a single pseudobosonic degree of freedom substantially reduces computational cost, as shown in Ref.~\cite{Larsson2022}.
	The pseudobosonic \gls{dmrg} ansatz remains numerically exact, achieving precision up to the controllable truncation error $\sqrt{\Delta}$; see~\cref{App:B}.
	\subsection{Projected-purified Pseudobosons}
	While the pseudoboson ansatz reduces the originally two-dimensional problem to an effective one-dimensional form in the compressed Fock basis, it comes at the price of two major shortcomings:
	First, for Hamiltonians with particle number conservation, the pseudoboson mapping and the subsequent compression evidently break the associated global $\mathrm{U}(1)$-symmetry, which is highly unfavorable for the following \gls{dmrg} calculation.
	Instead of many previously sparse tensors, it now has to deal with a large dense tensor, resulting in a significant slowdown of matrix operations.
	Secondly, since single-site \gls{dmrg}~\cite{Hubig2015,Hubig17} involves optimizing $\mathcal{O}(L\chi^3m)$ parameters, the polynomially large local dimension $m$ of the emerging pseudoboson sites drastically increases the computational complexity.
	To mitigate these issues we combine the pseudoboson mapping with the \gls{pp} method~\cite{Stolpp2021,Koehler2021}.
	The key idea of the PP method is to implement an artificial $\mathrm{U}(1)$-symmetry by first, doubling each local pseudoboson Hilbert space and then, enforcing that the occupation number on the bath site $\tilde{\Sigma}_B$ is constrained by the local gauge condition $g(\tilde{\Sigma}_P)= m-\tilde{\Sigma}_P$.
	Upon extending the original hard-core creation and annihilation operators as follows
	\begin{align}
		\hat{a}_{\omega}^{\dagger}\longrightarrow \hat{a}_{\scriptscriptstyle{P;\omega}}^{\dagger}\otimes \hat{\beta}_{\scriptscriptstyle{B;\omega}}^{\nodagger},\quad \hat{a}_{\omega}^{\nodagger}\longrightarrow \hat{a}_{\scriptscriptstyle{P;\omega}}^{\nodagger}\otimes \hat{\beta}^{\dagger}_{\scriptscriptstyle{B;\omega}}
	\end{align}
	the resulting Hamiltonian acting on the enlarged Hilbert space will preserve the sum of the number of physical particles and bath particles.
	On the \gls{mps} level, the mapping unfolds as follows:
	First, each site tensor $\mathbb{A}^{\tilde{\Sigma}_j}_{j}$ of \cref{eq:mps} is doubled, i.e., one obtains $\mathbb{A}^{\tilde{\Sigma}_{j;P},\tilde{\Sigma}_{j;B}}_{j;\chi_{j-1},\chi_j}$.
	Then, upon enforcing the gauge condition, the site tensors decompose under the artificial $\mathrm{U}(1)$ symmetry into tensor blocks
	\begin{align}
		T^{\scriptscriptstyle{\Sigma_{j;P}\Sigma_{j;B}}}_{\scriptscriptstyle{j;\alpha_{j-1}\alpha_j}}\delta(\Sigma_{\scriptscriptstyle{j;P}} + \Sigma_{\scriptscriptstyle{j;B}} + \alpha_{\scriptscriptstyle{j}} -\alpha_{\scriptscriptstyle{j-1}} )
	\end{align}
	where $\alpha_{j-1}$ and $\alpha_j$ label irreducible representations.
	This leads to the factorization of the physical and bath site tensors and we restore a block structure of sparse tensors labeled by the pseudoboson occupation quantum number $\tilde{\Sigma}$. 
	In~\cref{fig:overview}(c) the flow of the $\mathrm{U}(1)$ labels is shown.
	Beyond restoring $\mathrm{U}(1)$ symmetry, the \gls{pp} mapping can also remedy the increase in complexity that arises from large local dimensions.
	The Schmidt spectrum $\{\Lambda^{\Sigma_{j;P}}_{j,\tau}\}$ of the bipartition between a physical and bath site relates to the diagonal elements of the reduced cluster density matrix in the physical occupation basis via
	\begin{align}
		\sum_{\tau}\left( \Lambda^{(\Sigma_{j;P})}_{j,\tau}\right)^2=\rho_{\scriptscriptstyle{\Sigma_{j;P}},\scriptscriptstyle{\Sigma_{j;P}}}.
	\end{align}
	Note that the diagonal elements of the reduced density matrix encode the excitation probabilities of the $\Sigma_{j;P}$-th state.
	Consequently, the block-diagonal structure of the projected-purified tensors  allows for natural truncation of the large local dimension by discarding blocks with negligible spectral weight of the corresponding Fock state during a \gls{dmrg} sweep.
	Additionally, the decomposition of the large pseudoboson site-tensor in many independent sparse blocks yield a significant speed-up in practice as one can solve the optimization problem during a local \gls{dmrg} update in parallel.
	This will be discussed in more detail in the following sections.
	\subsection{Amenable Systems}\label{sec:Model}
	\label{subsection:am_systems}
	\begin{figure*}
		 \includegraphics{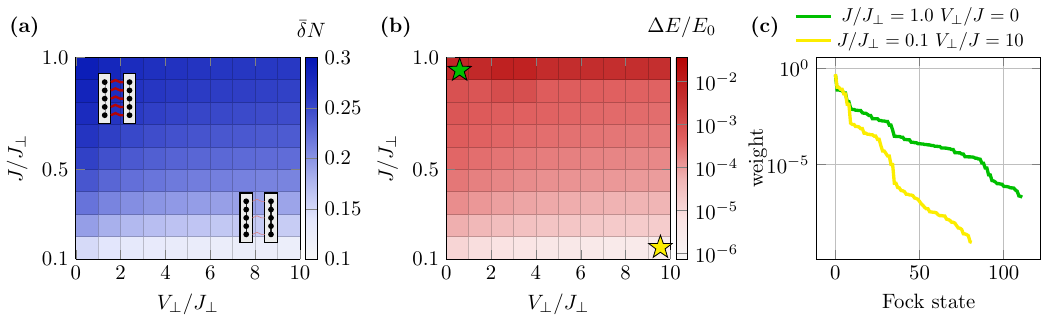}
		\caption{
			\textbf{Correlation between cluster particle-number fluctuations and ground-state accuracy}
			\textbf{(a)} Average cluster particle-number fluctuation $\bar{\delta} N$ versus $J/J_{\perp}$ and $V_{\perp}/J_{\perp}$. 
			\textbf{(b)} Relative ground-state energy error $\Delta E/E_0$ of the \gls{a3p-dmrg} for the same parameters. 
			Results are shown for an $L=W=8$ square lattice with open boundary conditions and $V/J_{\perp}=0$. 
			The effective basis dimension is fixed at $m=93$ ($N\leq3$).
			\textbf{(c)} Fock-basis state spectrum of the ground state for two special points (yellow and green star) in the $J/J_{\perp}-V_{\perp}/J_{\perp}$ plane.
		}
		\label{fig:fluctuation}
	\end{figure*}
	As discussed above, suppressed cluster particle number fluctuations $\delta N$ lead to an efficient, low-dimensional pseudobosonic representation for a fixed precision $\Delta$. 
	These fluctuations are controlled by the competition between intra-cluster and inter-cluster correlations.
	Microscopically, this interplay is directly related to the parameters of the underlying model.
	To illustrate this connection and identify the regimes where the pseudoboson ansatz excels or breaks down, we consider a simple yet generic model of hard-core bosons with nearest-neighbor repulsion on a square lattice. 
	The Hamiltonian for this system reads
	\begin{equation}
		\hat{H} = \hat{H}_{\parallel} + \hat{H}_{\perp} + \hat{H}_{\rm{int}} - \mu \sum_{j,\omega} \hat{n}_{j,\omega},
		\label{eq:H_total}
	\end{equation}
	with
	\begin{align}
		\hat{H}_{\perp} &= -J_{\perp} \sum_{j=1}^{L} \sum_{\omega=1}^{W} 
		\left( \, \hat{a}_{j,\omega}^\dagger \hat{a}_{j,\omega+1}^{\nodagger} + \mathrm{h.c.} \right), 
		\label{eq:H_perp}
		\\[0.2cm]
		\hat{H}_{\parallel} &= -  J\sum_{j=1}^{L-1} \sum_{\omega=1}^{W}
		\left( \hat{a}_{j,\omega}^\dagger \hat{a}_{j+1,\omega}^{\nodagger} + \mathrm{h.c.} \right),
		\label{eq:H_parallel}
		\\[0.2cm]
		\hat{H}_{\rm{int}} &=  V_{\perp}\sum_{j,\omega} \, \hat{n}_{j,\omega} \hat{n}_{j,\omega+1} 
		+  V\sum_{j,\omega} \hat{n}_{j,\omega} \hat{n}_{j+1,\omega}.
		\label{eq:H_int}
	\end{align}
	$\hat{H}_{\perp}$ describes intra-cluster hopping with amplitude $J_{\perp}$ while $\hat{H}_{\parallel}$ accounts for the hopping between clusters with amplitude $J$.
	Nearest-neighbor density-density interactions along and across clusters are included in $\hat{H}_{\rm{int}}$, with couplings $V$ and $V_{\perp}$, respectively. 
	Finally, $\mu$ is the chemical potential that controls the average number of particles.
	We tune both the hopping amplitudes and interaction strengths to understand how the Hamiltonian parameters influence $\delta N$ and the effectiveness of the pseudoboson ansatz.
	\cref{fig:fluctuation}(a) shows the average fluctuation in the cluster-particle number for a system of size \(L=W=8\) with open boundary conditions and $n=\nicefrac{N}{\mathcal{L}}=\nicefrac{1}{2}$.
	As expected, $\delta N$ decreases systematically when inter-cluster hopping is reduced $(J \ll J_{\perp})$, while the reduction due to increasing intra-cluster interaction $V_{\perp}$ is less pronounced.
	Because increasing $V_{\perp}$ increases the energy cost $\Delta\mu(N_j)$ for adding or removing a particle in the cluster, the weak-coupling scaling $\delta N \sim J^2 / |\Delta\mu(N_j)|^2$ (see~\cref{App:C}) accurately reproduces the observed reduction in cluster fluctuations.
	A small $\delta N$ restricts the many-body Hilbert space effectively: higher-occupancy cluster sectors are suppressed, and entanglement between clusters is reduced.
	For a fixed local basis size $m$, this leads to higher numerical precision, as illustrated in \cref{fig:fluctuation}(b).
	Conversely, finite inter-cluster interactions $(V>0)$ introduce correlations between pseudobosons, broadening the effective Hilbert space and requiring a larger basis to maintain accuracy.
	For $J/J_{\perp}=0.1$, we revisit the scenario discussed in~\cref{subsec:bennett_bound}, $(\bar{N}=\nicefrac{1}{2},\delta N=\nicefrac{1}{10})$. 
	Importantly, although the effective basis truncation to $N\leq 3$ is more restrictive than the bound suggested by the Bennett criterion ($N\leq 5$), we nevertheless observe a relative ground-state energy error $\Delta E/E_0$ well below the target threshold $\Delta$.
	\cref{fig:fluctuation}(c) further illustrates how different parameter regimes give rise to markedly distinct structures of the ground state.
	For $J/J_{\perp}=0.1$ and $V_{\perp}/J_{\perp}=10$, the weights of the Fock-basis states decay rapidly, reflecting the strong suppression of higher-occupancy sectors.
	In contrast, in the opposite limit $J/J_{\perp}=1.0$ and $V_{\perp}/J_{\perp}=0$, the decay is significantly slower, indicating a much broader distribution over Fock-state configurations.
	This analysis highlights that dilute, low-variance regimes 
	with weak inter-cluster correlations -- directly controlled by the Hamiltonian parameters --
	are particularly well suited to the pseudoboson approach introduced here, even when the Bennett criterion is not strictly saturated.
	While the model is not intended to capture complex many-body physics, it provides a useful proxy for experimentally relevant systems.
	Cold-atom experiments naturally operate in low-filling, weak-tunneling regimes, particularly in deep optical lattices with synthetic dimensions, where particles are localized to individual sites and intra-cluster hopping and interactions are encoded in the internal atomic structure~\cite{Boada2012,Celi2014}.
	Such setups allow the realization of interacting models in dilute, weakly coupled regimes, even in the presence of an artificial magnetic field~\cite{Gross2017,Altman2021,Goldman2016,Aidelsburger2013}, which increases frustration by flattening the single-particle spectrum and provides an additional tuning parameter to suppress cluster fluctuations.
	In this context, weakly coupled $W$-legged flux ladders~\cite{Lehur2015,Buser2020,Palm2022} offer an experimentally relevant platform for probing the stability of topological order~\cite{Tai2017,Impertro2025}.
	Similarly, comb-like geometries of dissipative hard-core bosons ($J\rightarrow 0$ on all but one site) can host transverse quantum fluids and potentially exhibit superfluid-to-supersolid transitions~\cite{Kuklov2024,Pollet2018}, although the persistence of such features in the dilute, weak-coupling limit remains uncertain.
\section{Adaptive 3P-DMRG}
\label{sec:MF}
While the pseudoboson mapping provides a natural mechanism to truncate particle-number sectors on cluster sites with small particle-number fluctuations consistent with the probabilistic bound derived above, the observation that weak inter-cluster coupling suppresses fluctuations (see~\cref{App:C}) suggests that further compression of the local basis within each degenerate particle-number sector is possible. 
To achieve this, we introduce a self-consistent mean-field procedure, conceptually related to density-matrix embedding theory~\cite{Reinhard2019}, which constructs an \textit{optimized} pseudoboson basis adapted to a given parameter regime.
In this picture, a single cluster is treated as an impurity coupled to a static mean-field bath that represents its neighboring sites, see~\cref{fig:overview}(d).
\begin{figure*}[t]
	 \includegraphics{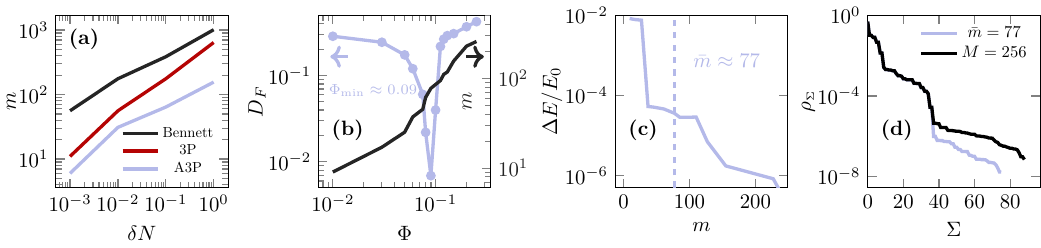}
	\caption{
		\textbf{Optimization of the effective pseudoboson basis}
		\textbf{(a)} Effective basis dimension $m$ versus particle-number fluctuation $\delta N$ for the Bennett bound, \gls{3p-dmrg}, and \gls{a3p-dmrg} approaches ($W=10$). 
		\textbf{(b)} Cost function $D_F$ (blue line, left axis) and effective basis dimension $m$ (black line, right axis) versus mean-field coupling $\Phi$.
		\textbf{(c)} Relative ground-state error versus effective basis dimension $m$.
		The dashed line marks the approximate minimum of $D_F$.
		\textbf{(d)} Spectrum of the averaged cluster reduced density matrix in the pseudoboson basis for the optimal and full bases. 
		Results are for an $L=W=8$ square lattice with open boundaries, $N=4$, and $V_{\perp}/J_{\perp}=V/J_{\perp}=0$, with discarded weight $\Delta=10^{-5}$.
	}
	\label{fig:basisdim}
\end{figure*}
Let $\hat{H}_{{\mathrm{c}}}$ denote the Hamiltonian of an isolated cluster, including kinetic terms $J_{\perp}$, intra-cluster interactions $V_{\perp}$ and effective chemical potential $\mu_{\mathrm{eff}}(\bar{N})$. 
The impurity site is coupled to its environment via
\begin{equation}
    \hat{H}_{\mathrm{env}} = - J \sum_{\omega} 
    \big( \hat{a}_{\omega}^\dagger \Phi_{\omega}^{\nodagger} + \Phi_{\omega}^\dagger \hat{a}_{\omega}^{\nodagger} \big)
    + V \sum_{\omega} \hat{n}_{\omega}^{\nodagger} \, \bar{n}_{\omega}^{\mathrm{(env)}},
\end{equation}
where $\Phi_{\omega}^{\nodagger}$ and $\bar{n}_{\omega}^{\mathrm{(env)}}$ are the mean-field parameters that represent the effect of the bath on the site $\omega$ of the cluster.
Within the mean-field approximation, the impurity Hamiltonian reads
\begin{align}
    \hat{H}_{\mathrm{imp}}\left(\{\Phi_{\omega}\},\{\bar{n}_{\omega}\}\right)
    = \hat{H}_{\mathrm{c}} + \hat{H}_{\mathrm{env}}\left(\{\Phi_{\omega}\},\{\bar{n}_{\omega}\}\right).
\end{align}
In the following, we further simplify the ansatz by assuming spatially homogeneous couplings $\Phi_{\omega}=\Phi$ and $\bar{n}_{\omega}=\bar{n}$ for all $\omega$.
Starting from an initial input $\Phi^{(0)}$, $\bar{n}^{(0)}$, the mean-field loop consists of the following steps:
\begin{enumerate}
    \item Diagonalize $\hat{H}_{\mathrm{imp}}[\Phi^{(k)},\bar{n}^{(k)}]$ to obtain the ground state
          $|\psi^{(k)}\rangle$.
    \item Update the mean-field parameters:
\[
\begin{aligned}
    \hspace{1cm} 
    \Phi^{(k+1)} &= \langle \psi^{(k)} | \sum_{\omega}\hat{a}_{\omega} | \psi^{(k)} \rangle, \\
    \hspace{1cm} 
    \bar{n}^{(k+1)} &= \langle \psi^{(k)} | \sum_{\omega}\hat{n}_{\omega} | \psi^{(k)} \rangle.
\end{aligned}
\]
    \item Optionally, mix old and new values to stabilize convergence:

            \[
\begin{aligned}
    \Phi^{(k+1)} \leftarrow (1-\eta)\Phi^{(k)} + \eta \Phi^{(k+1)},
          \ \rm with \ 0<\eta\le 0.5.
\end{aligned}
\]
\end{enumerate}
The procedure is repeated until $|\Phi^{(k+1)}_{\omega}-\Phi^{(k)}_{\omega}|<\varepsilon$ and
$|\bar{n}^{(k+1)}_{\omega}-\bar{n}^{(k)}_{\omega}|<\varepsilon$ for all $\omega$.
Once the self-consistency loop has converged, the ground state $|\psi_{\mathrm{MF}}\rangle$ 
of the mean-field Hamiltonian provides access to the spectral weights $\lambda^{\scriptscriptstyle{\Sigma}}_{\scriptscriptstyle{\rm MF}}$ of each pseudobosonic basis state $\ket{\Sigma}$ via
\begin{align}
\lambda^{\scriptscriptstyle{\Sigma}}_{\scriptscriptstyle{\rm MF}} = |\braket{\Sigma | \psi_{\scriptscriptstyle{\rm MF}}}|^2.
\end{align}
The spectrum $\{\lambda^{\scriptscriptstyle{\Sigma}}_{\scriptscriptstyle{\rm MF}}\}_{\scriptscriptstyle{\Sigma \in \mathcal{M}}}$ corresponds to the eigenvalues of the cluster reduced density matrix $\hat{\rho}_{\scriptscriptstyle{\rm MF}}$ in the Fock basis.
Fixing a discarded probability mass $\Delta$, we retain only the $\bar{m}$ most significant basis states whose cumulative weight is $1 - \Delta$.
This defines the optimized pseudoboson basis
\begin{align}
\{\ket{\phi_1},\dots,\ket{\phi_{\bar{m}}}\}
\equiv 
\{\ket{\Sigma}:\lambda^{\Sigma}_{\scriptscriptstyle{\rm MF}}\ge\varepsilon\},
\end{align}
where $\varepsilon$ is the truncation threshold corresponding to the discarded weight $\Delta$.
This reduced set spans an effective low-rank subspace for the \gls{a3p-dmrg} calculation.
In~\cref{fig:basisdim}(a) we demonstrate how the mean-field optimization significantly reduces the effective dimensionality of the cluster Hilbert space, i.e., $\bar{m} < m$.
The basis dimension $m$ depends monotonically on the mean-field parameter $\Phi$, thereby constituting a variational ansatz of controllable precision, see~\cref{fig:basisdim}(b).
Finally, the full two-dimensional Hamiltonian is projected cluster by cluster onto the effective low-rank submanifold spanned by the optimized pseudoboson basis
\begin{align}
    \hat{H}_{j,\mathrm{eff}}=\hat{\mathcal{P}}^{\dagger}\hat{H}_j\hat{\mathcal{P}}^{\nodagger}
\end{align}
where $\hat{\mathcal{P}}=\left[ \lvert \phi_1 \rangle ,..., \lvert \phi_{\bar{m}} \rangle\right]\in \mathbb{C}^{M\times \bar{m}}$ and $\hat{H}_{ j,\rm eff}\in \mathbb{C}^{\bar{m}\times \bar{m}}$.
The full projected Hamiltonian $\hat H_{\rm eff}$ is then obtained by assembling the cluster-projected operators and their inter-cluster couplings.
Therefrom, the ground state $\ket{\Psi}$ of $\hat{H}_{\rm eff}$ is computed using \gls{3p-dmrg} on this subspace.
A natural measure of how well the effective mean-field basis captures the true two-dimensional correlations is given by the distance between the averaged cluster reduced density matrix of the DMRG ground state, $\bar{\hat{\rho}}^{(\varepsilon)}_{\scriptscriptstyle{\rm DMRG}} = L^{-1}\sum_{j=1}^{L}\hat{\rho}_{\scriptscriptstyle{\mathrm{DMRG}},j}^{(\varepsilon)}$, and the truncated mean-field density matrix $\hat{\rho}^{(\varepsilon)}_{\scriptscriptstyle{\rm MF}}$, evaluated using the Frobenius norm:
\begin{align} \label{eq:sc}
D_F =|| \hat{\rho}^{(\varepsilon)}_{\scriptscriptstyle{\rm MF}} -\bar{\hat{\rho}}^{(\varepsilon)}_{\scriptscriptstyle{\rm DMRG}}||_F^2.
\end{align}
~\cref{eq:sc} thus serves as a high-level self-consistency condition in the conventional mean-field sense, quantifying how accurately the optimized pseudoboson basis captures the entanglement structure of the full system.
Importantly, it can be shown that $D_F$ in the vicinity of the minimum is locally convex as a function of the mean-field coupling $\Phi$.
Hence, once the loop has converged, if $D_F$ is of order $\sqrt{\Delta}$, the obtained basis can be regarded as \textit{optimal}, efficiently encoding the relevant two-dimensional correlations (see~\cref{App:D}).
In~\cref{fig:basisdim}(b), the minimization of $D_F$ as a function of $\Phi$ is shown for an open boundary square lattice of size $L=W=8$ with $N=4$ particles, $J/J_{\perp}=0.1$ and $V_{\perp}/J_{\perp}=0$, $V/J_{\perp}=0$.
~\cref{fig:basisdim}(c) shows the relative error in the ground-state energy of the \gls{a3p-dmrg} approach versus the effective pseudobosons basis dimension during the mean-field loop, ~\cref{fig:basisdim}(d) shows the spectrum of the density matrix of the ground state at optimal basis size and at full basis size.
This adaptive mean-field optimization captures the correlations between each cluster and its environment at modest computational cost. 
It is particularly effective in regimes with weak inter-cluster coupling $J$ 
and small density fluctuations, where the mean-field bath approximation is accurate. 
Systematic improvements could be achieved by extending the impurity to multiple clusters 
or by introducing additional mean-field parameters, such as bond currents in flux phases 
or site-dependent $\Phi_{\omega}$ and $\bar{n}_{\omega}$.

\section{Benchmarks}\label{sec:Benchmarks}
In this section, we benchmark the \gls{a3p-dmrg} against conventional hard-core bosonic \gls{dmrg} to assess its efficiency and scalability for two-dimensional dilute systems.
Building on the hard-core boson model introduced in Sec.~\ref{sec:Method}, we extend the Hamiltonian by incorporating a magnetic field through complex hopping amplitudes, which is both experimentally relevant and particularly well suited to the pseudoboson formulation, while posing a significant challenge for methods affected by the sign problem~\cite{Beach2008}.
\begin{figure}[b]
	 \includegraphics{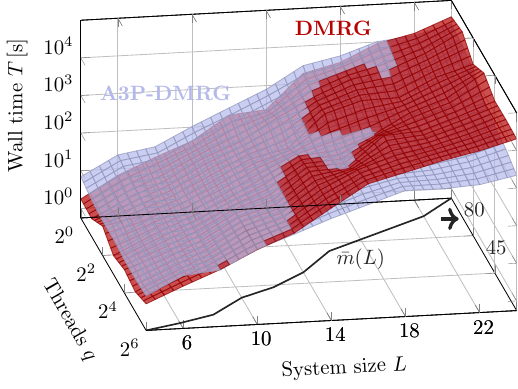}
	\caption{
		\textbf{Runtime scaling of \gls{dmrg} with hard-core boson and optimized pseudoboson bases}
		Wall time $T$ of \gls{dmrg} runs as a function of system length $L$ and number of threads (logical cores) for the hard-core boson basis (red plane) and the optimized pseudoboson basis (blue plane).
		All calculations were performed on the same computing node for a square lattice with open boundary conditions, density $n=1/L$, weak inter-cluster coupling $J/J_{\perp}=0.1$, vanishing interactions $V=V_{\perp}=0$, and flux per plaquette $\alpha=\nicefrac{1}{4}$. 
		The \gls{dmrg} staging parameters (number of sweeps, bond dimension, and convergence criteria) were kept identical for both bases.  
		The secondary axis in the $x$–$y$ plane indicates the effective basis dimension $\bar{m}(L)$ as a function of system size.
	}
	\label{fig:runtimevsthreads}
\end{figure}
\begin{figure*}
	\centering
	 \includegraphics{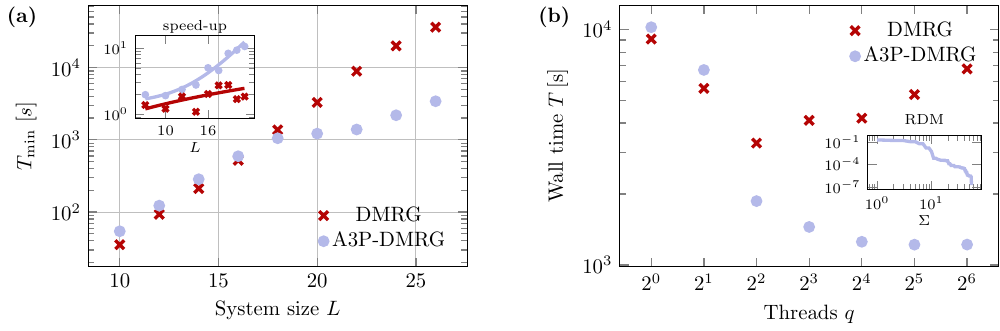}
	\caption{
		\textbf{Summary of runtime scaling and multi-threading performance of \gls{dmrg} with hard-core and pseudoboson bases}
		\textbf{(a)} Wall time of \gls{dmrg} runs as a function of system length $L$ for the hard-core boson basis (red crosses) and the optimized pseudoboson basis (light blue circles), evaluated at the respective optimal thread count ($T_{\mathrm{min}}$).
		The inset shows the multi-threading speed-up factor $T_{\mathrm{min}}/T_1$ as a function of the system length $L$.
		\textbf{(b)} Wall time T of a \gls{dmrg} run as a function of the number of threads used for hard-core bosons (red crosses) and pseudobosons (light blue circles).
		Calculations were performed on the same computing node using open boundary conditions on a square lattice of size $L=20$ with density $n = 1/L$, weak inter-cluster coupling $J/J_{\perp} = 0.1$, $V = V_{\perp} = 0$ and $\alpha=\nicefrac{1}{4}$.
		The inset shows the average cluster reduced density matrix of the ground state.
	}
	\label{fig:runtime_scaling}
\end{figure*}
In Landau gauge, we obtain the extended Hamiltonian by modifying $\hat{H}_{\perp}$ as follows
\begin{align}
    \hat{H}_{\perp} = -J_{\perp}\sum_{j=1}^{L} \sum_{\omega=1}^{W} 
\left(e^{i \phi_{\omega}} \, \hat{a}_{j,\omega}^\dagger \hat{a}_{j,\omega+1}^{\nodagger} + \mathrm{h.c.} \right)
\end{align}
where $\phi_{\omega}=2\pi\alpha\omega$ is the Peierls phase induced by the finite magnetic flux through the lattice.
In the subsequent simulations, we focus on the low-filling regime $n\approx 1/L$ and weak inter-cluster coupling $J/J_{\perp}\approx 0.1$, where the approach is especially effective.
We furthermore fix the flux per plaquette value to $\alpha=\nicefrac{1}{4}$ which introduces substantial band flattening.
All simulations are performed using identical DMRG staging parameters and equivalent computational resources to ensure a fair comparison.
The main results are twofold.
On the one hand, we demonstrate that the sparse projected–purified tensor structure emerging in the pseudoboson basis enables efficient multithreading parallelization, exhibiting a superlinear power-law scaling with system size in contrast to the linear scaling observed for the conventional hard-core ansatz.
On the other hand, we show that our ansatz improves the time-to-result complexity for large, dilute systems.
We first examine the systematic runtime scaling with system size, emphasizing the effect of multithreading parallelization; see~\cref{fig:runtimevsthreads}.  
The runtime of the \gls{a3p-dmrg} decreases consistently with the number of threads, whereas the scaling behavior in the original hard-core boson basis is less regular and often non-monotonic.
Although the wall time $T$ of the hard-core bosonic \gls{dmrg} remains smaller for system sizes up to $L \le 16$, this trend reverses beyond $L\!\approx\!16$, where the pseudoboson method achieves a significantly faster time-to-result. 
In general, the wall time $T$ of the hard-core formulation grows exponentially with system size $L$, whereas the \gls{a3p-dmrg} shows exponential scaling only up to $L\!\approx\!14$. 
For larger system sizes approaching the dilute limit, we observe a pronounced saturation, see~\cref{fig:runtime_scaling}(a), signaling the onset of a runtime plateau in the large-$L$, low-filling regime.
The dimension of the effective pseudobosonic basis $\bar{m}$ as a function of the system size $L$ is shown as an inset in~\cref{fig:runtimevsthreads} and is chosen across all $L$ consistently with a discarded probability mass of $\Delta=10^{-5}$.
The crossover in efficiency for larger systems can be attributed to several factors. 
First, as the system grows, the effective pseudoboson basis dimension $\bar{m}$ increases, resulting in a larger number of independently optimizable tensor blocks in the sparse projected–purified representation (see the inset of~\cref{fig:runtime_scaling}(a)).
The \gls{a3p-dmrg} (blue circles) exhibits a power-law speedup (blue line, fit) $\nicefrac{T_{\mathrm{min}}}{T_1}\sim L^{\gamma}$ with $\gamma\approx3.13$, whereas the hard-core boson \gls{dmrg} (red crosses) scales linearly ($\gamma \approx 1$) (red line, fit).
Here, $T_{\mathrm{min}}$ denotes the minimal wall time obtained at the optimal number of threads $q_{\mathrm{opt}}$ minimizing $T_q(L)$, with $q \in \{1, 2, 4, 8, 16, 32, 64\}$, and $T_1$ the single-thread runtime.
The enhanced parallel speed-up is directly linked to the spectrum of the cluster reduced density matrix $\rho_{\scriptscriptstyle{\Sigma}}$, whose number of significant modes determines the ideal thread count for efficient blockwise optimization; see~\cref{fig:runtime_scaling}(b).
Importantly, also without any parallelization efforts, the pseudoboson approach produces faster time-to-result calculations for $L \geq 20$.
Secondly, the adaptive basis reduction effectively biases the optimization toward the physically relevant low-filling manifold, thereby improving convergence in a manner similar to providing a preconditioned initial state.
Taken together, these results demonstrate that the \gls{a3p-dmrg} formulation substantially enhances parallel efficiency and improves scaling with system size, enabling simulations of large two-dimensional dilute quantum systems beyond the practical reach of conventional \gls{dmrg} approaches.
\section{Conclusion}\label{sec:Conclusion}
In this work, we introduced the \textit{adaptive pseudoboson DMRG} (\gls{a3p-dmrg}) method and established it as a versatile and efficient approach to obtain ground states of dilute, hard-core bosonic two-dimensional systems. 
We demonstrated that simple \textit{a priori} probabilistic arguments can drastically reduce the effective dimensionality of the many-body Hilbert space by identifying the physically relevant particle-number sectors. 
To exploit this intrinsic bias, we introduced the concept of \textit{pseudobosons} as collective many-body degrees of freedom associated with Fock-basis configurations. 
This reformulation enables an adaptive, physically motivated truncation of the local basis and yields a substantial acceleration in \gls{dmrg} simulations for large systems with a controllable accuracy.
Given its compatibility with existing tensor-network infrastructures such as time evolution methods~\cite{Paeckel2019} and its natural amenability to parallelization, the \gls{a3p-dmrg} provides a promising route for the large-scale numerical exploration of dilute regimes in a variety of experimentally relevant platforms, including ultracold atomic lattices, moir\'e excitons, and photonic quantum fluids.
Importantly, the method is not restricted to particle statistics but can be extended to other $\mathrm{U}(1)$ quantum numbers such as momentum, spin or more abstract conserved quantities.
Since the approach is particularly effective when only a few $\mathrm{U}(1)$ sectors carry significant spectral weight, systems that exhibit strong delocalization and broad particle-number distributions may still be efficiently represented in a dual (here momentum) basis.
Consequently, even in regimes where the pseudoboson method becomes inefficient in one representation, there often exists a complementary basis in which it can recover its advantages.
Other natural near-term extensions include fermionic and spinful systems.
 These extensions would enable us to investigate parameter regimes, namely low filling, of inhomogeneous $t-J$ planes, only connected via spin-spin coupling, in the search for high $T_c$ superconductivity,which are not accessible using approaches such as \gls{mps} combined with mean-field theory~\cite{Bollmark2023,Koehler2025}.
\acknowledgments
    \ 
    The authors would like to thank Martin Grundner, Felix Palm and Adrian Kantian for fruitful discussions.
    FJP and SP acknowledge support by the Deutsche Forschungsgemeinschaft (DFG, German Research Foun- dation) under Germany’s Excellence Strategy-426 EXC- 2111-390814868.
    TK acknowledge support by the United Kingdom’s Engineering and Physical Sciences Research Council (Grant No. EP/W022982/1).
    All calculations were performed using the \textsc{SyTen}-toolkit~\cite{SyTen,Hubig17}.

\onecolumngrid
\begin{appendix}
\section{Commutation Relation}\label{App:A}
Here, we demonstrate that the pseudoboson representation preserves the hard-core bosonic algebra, i.e., it satisfies the commutation relations given in~\cref{eq:ccr}.
Owing to the orthogonality of the basis states, the first two conditions follow straightforwardly.
It therefore suffices to explicitly evaluate the remaining commutator, which we compute below.
\begin{align}
	\{\hat{a}_{\omega},\hat{a}_{\omega}^{\dagger}\} &= \sum_{\scriptscriptstyle{\Sigma,\Sigma^{\prime}} \in \mathcal{B}^{\complement}_{\omega}} \left( \left(\ket{\Sigma}\bra{\Sigma + 2^{{\omega}-1}}\right) \left(\ket{\Sigma^{\prime}}\bra{\Sigma^{\prime} + 2^{{\omega}-1}}\right)^{\dagger} + \left(\ket{\Sigma^{\prime}}\bra{\Sigma^{\prime}  + 2^{{\omega}-1}}\right)^{\dagger}\left(\ket{\Sigma}\bra{\Sigma + 2^{{\omega}-1}}\right) \right) \nonumber \\
	&=  \sum_{\scriptscriptstyle{\Sigma} \in \mathcal{B}^{\complement}_{\omega}}  \left( \ket{\Sigma}\bra{\Sigma} +  \ket{\Sigma +2^{{\omega}-1}}\bra{\Sigma+2^{{\omega}-1}} \right). \nonumber
\end{align}
Since, if $\Sigma \in \mathcal{B}^{\complement}_{\omega} \Rightarrow \Sigma +2^{{\omega}-1} \equiv \Lambda \in \mathcal{M} \setminus \mathcal{B}^{\complement}_{\omega} = \mathcal{B}_{\omega}$, it follows that
\begin{equation}
	\{\hat{a}_{\omega},\hat{a}_{\omega}^{\dagger}\} = \sum_{\scriptscriptstyle{\Sigma} \in \mathcal{B}^{\complement}_{\omega}} \ket{\Sigma}\bra{\Sigma} + \sum_{\scriptscriptstyle{\Lambda} \in \mathcal{B}_{\omega}} \ket{\Lambda}\bra{\Lambda}= \sum_{\scriptscriptstyle{\Sigma} = 0}^{M-1} \ket{\Sigma}\bra{\Sigma} = \mathbb{I}.
\end{equation}
\section{Pseudobosons for soft-core bosons and fermions}\label{App:B2}
The pseudoboson mapping for hard-core bosons can be generalized to fermionic and soft-core bosonic degrees of freedom in a straightforward manner.
While the many-body basis of a fermionic extension of the mapping is identical, the local operator algebra needs to be modified such that it respects the fermionic sign.
On the pseudoboson level, assuming normal ordering of lattice sites, one defines a parity operator counting all fermions on cluster sites with index $j^{\prime}<j$
\begin{align}
    \hat{P}_{j^{\prime}< j}= \prod_{j'<j} (-1)^{\hat N_{j'}}.
\end{align}
Additionally, to obtain the correct sign for the matrix elements of the local fermionic operators one needs to account for the fermions on the $j$th cluster site itself.
We define the on-site sign function as
\begin{align}
    s(\Sigma_j,\omega)=(-1)^{\sum_{\omega^{\prime}< \omega}\sigma_{j,\omega^{\prime}}(\Sigma_j)}  
\end{align}
where $\sigma_{j,\omega^{\prime}}(\Sigma_j)$ denotes the inverse mapping from the pseudoboson labels to the corresponding hard-core boson occupations, as defined in~\cref{eq:pbmapping}.
With this definition, the fermionic creation (annihilation) operator $f^{(\dagger)}_{j,\omega}$ reads 
\begin{align}
    f^{(\dagger)}_{j,\omega}= \hat{P}_{j^{\prime}< j}\sum_{\Sigma_j \in \mathcal{B}_{\omega}} s(\Sigma_j,\omega)\left( \ket{\Sigma_j - 2^{\omega-1}}\bra{\Sigma_j} \right)^{(\dagger)}.
\end{align}
The mapping generalizes to soft-core bosons with arbitrary maximal on-site occupation $d$ by employing a $(d+1)$-ary encoding of the cluster Fock states yielding
\begin{align}
    \Sigma_j = \sum_{\scriptscriptstyle{\omega=1}}^{W}\sigma_{j,\omega}(d+1)^{\omega-1},
\end{align}
where $\sigma_{j,\omega}=0,1,...,d$.
The matrix elements of the soft-core bosonic creation (annihilation) operators $\hat{b}^{(\dagger)}_{j,\omega}$ are given by
\begin{align}
    \hat{b}_{j,\omega} &= \sum_{n=1}^{d}\sum_{\Sigma \in \mathcal{B}_{\omega}^{(n)}}\sqrt{n}\ket{\Sigma-(d+1)^{\omega -1}}\bra{\Sigma}\\
    \hat{b}^{\dagger}_{j,\omega} &= \sum_{n=1}^{d}\sum_{\Sigma \in \mathcal{B}_{\omega}^{(n+1)}}\sqrt{n+1}\ket{\Sigma}\bra{\Sigma-(d+1)^{\omega -1}}
\end{align}
where $\mathcal{B}^{(n)}_{\omega}:=\{ \Sigma \in \mathcal{M} \vert \sigma_{\omega}(\Sigma)=n \}$.
\section{Pseudoboson error estimation}\label{App:B}
For a cluster of $W$ sites we can express any pure state in its pseudoboson representation
\begin{align}
    \ket{\Psi}=\sum_{\scriptscriptstyle{\Sigma}=0}^{M-1} \sqrt{\lambda_{\scriptscriptstyle{\Sigma}}} \ket{\Sigma}.
\end{align}
Suppose we keep only the first $m$ basis states such that $\sum_{\Sigma=0}^m|\lambda_{\scriptscriptstyle{\Sigma}}|\le 1-\Delta$ with $\Delta>0$.
The state $\ket{\Psi}$ in the truncated basis then reads
\begin{align}
    \ket{\Psi_{\scriptscriptstyle{\mathrm{trunc}}}}=\frac{\hat{\mathcal{P}}\ket{\Psi}}{|| \hat{\mathcal{P}}\ket{\Psi} ||}, \ \hat{\mathcal{P}}=\sum_{\scriptscriptstyle{\Sigma}=0}^m \ket{\Sigma}\bra{\Sigma}.
\end{align}
Since $|\braket{\Psi|\Psi_{\scriptscriptstyle{\mathrm{trunc}}}}|^2=1-\Delta$ and $\Delta \ll1$, we can estimate the error of the truncation as
\begin{align}
    ||\ket{\Psi}-\ket{\Psi_{\scriptscriptstyle{\mathrm{trunc}}}}||=\sqrt{2(1-\sqrt{1-\Delta})}\approx \sqrt{\Delta}.
\end{align}
For an arbitrary well-defined bounded observable $\hat{O}$ with operator norm $||\hat{O}||$ the error in the expectation values introduced by the truncation obeys
\begin{align}
    |\braket{\Psi|\hat{O}|\Psi}-\braket{\Psi_{\scriptscriptstyle{\mathrm{trunc}}}|\hat{O}|\Psi_{\scriptscriptstyle{\mathrm{trunc}}}}|\le 2 ||\hat{O}||\sqrt{\Delta}.
\end{align}
Generally, we expect an error of the order $\sqrt{\Delta}$ for the pseudoboson method in all relevant observables.
\section{Weak Coupling scaling of fluctuations}\label{App:C}
Weak inter-cluster hopping combined with finite intra-cluster interactions significantly suppress particle-number fluctuations.
Below, we briefly outline how this behavior can be expressed formally.
Consider a generic Hamiltonian of the form
\begin{align}
    \hat{H}=\sum_{j=1}^{L}\hat{H}_{\perp,j}-J\sum_{\braket{i,j}}\sum_{\omega=1}^{W}\Big( \hat{a}^{\dagger}_{i,\omega}\hat{a}_{j,\omega} +\mathrm{h.c.} \Big).
\end{align}
If $J=0$, then the cluster particle number operator $\hat{N}_j=\sum_{\omega=1}^{W}\hat{n}_{j,\omega}$ is a conserved observable, i.e.,
\begin{align}
    \big[ \hat{H},\hat{N}_j \big]=0
\end{align}
meaning that there is also no particle number fluctuation $\delta N_j=0$.
Tuning $0<J\ll1$, by first order perturbation theory the corrections are transition elements between adjacent particle number sectors $N_j\rightarrow N_j\pm1$ of the form
\begin{align}
t^{(\tau)}_{\scriptscriptstyle{N_j\rightarrow \pm 1}} = \frac{\braket{E^{(\tau)}_{\scriptscriptstyle{N_j\pm1}}|J\sum_{\braket{i,j}}\sum_{\omega}( \hat{a}^{\dagger}_{i,\omega}\hat{a}_{j,\omega} +\mathrm{h.c.} )|E_{\scriptscriptstyle{N_j}}}}{E_{\scriptscriptstyle{N_j\pm1}}-E_{\scriptscriptstyle{N_j}}} \sim \frac{J}{\Delta\mu(N_j)},
\end{align}
where $\Delta\mu(N_j)=E_{\scriptscriptstyle{N_j\pm1}}-E_{\scriptscriptstyle{N_j}}$ is the cluster charge gap.
The fluctuation can be related to the sum of the probabilities of leakages to adjacent particle number sectors which are directly related to the transition elements $t^{(\tau)}_{\scriptscriptstyle{N_j\rightarrow \pm 1}}$ via 
\begin{align}
    \delta N_j\sim\sum_{\tau}p^{(\tau)}_{\scriptscriptstyle{N_j\pm1}}=\sum_{\tau}\Big|t^{(\tau)}_{\scriptscriptstyle{N_j\rightarrow \pm 1}}\Big|^2\sim \mathcal{O}\left( \frac{|J|^2}{|\Delta\mu(N_j)|^2} \right).
\end{align}
\section{Properties of \texorpdfstring{$D_F$}{D\_F}}
\label{App:D}
To assess how accurately the mean-field truncation captures the underlying two-dimensional physics, we seek an upper bound on the minimum of the \textit{cost function} $D_F$, defined in~\cref{eq:sc}, and to establish its local convexity around the minimum as a function of the mean-field parameter $\Phi$.

We begin by deriving a bound on $D_F$,
\begin{align}\label{eq:appbound}
    \sqrt{D_F}
    = \bigl\| \rho^{(\varepsilon)}_{\scriptscriptstyle{\rm MF}} - \bar{\rho}^{(\varepsilon)}_{\scriptscriptstyle{\rm DMRG}} \bigr\|_F
    \leq 
    \bigl\| \rho^{(\varepsilon)}_{\scriptscriptstyle{\rm MF}} - \rho_{\scriptscriptstyle{\rm MF}} \bigr\|_F
    +
    \bigl\| \rho_{\scriptscriptstyle{\rm MF}} - \bar{\rho}^{(\varepsilon)}_{\scriptscriptstyle{\rm DMRG}} \bigr\|_F .
\end{align}
The Frobenius norm of a linear map $A \in \mathbb{R}^{m\times n}$ is defined as
\begin{align}
    \|A\|_F
    = \sqrt{\mathrm{tr}\{AA^{\dagger}\}}
    = \sqrt{\sum_{k=1}^m \sum_{l=1}^n |a_{k,l}|^2 } .
\end{align}
The first term on the right-hand side of~\cref{eq:appbound} corresponds to the truncation error and is given by the sum of squared discarded spectral weights $\lambda_{\scriptscriptstyle{\Sigma}}$,
\begin{align}
    \bigl\| \rho^{(\varepsilon)}_{\scriptscriptstyle{\rm MF}} - \rho_{\scriptscriptstyle{\rm MF}} \bigr\|_F
    = \sqrt{\sum_{\scriptscriptstyle{\Sigma}\in\mathcal{S}} \lambda_{\scriptscriptstyle{\Sigma}}^2}
    \leq \left(\varepsilon \underbrace{\sum_{\scriptscriptstyle{\Sigma}\in\mathcal{S}} \lambda_{\scriptscriptstyle{\Sigma}}}_{\Delta}\right)^{\frac{1}{2}}
    = \sqrt{\varepsilon \Delta} ,
\end{align}
where $\mathcal{S}$ denotes the set of discarded pseudobosonic basis labels.
The second term, $\bigl\| \rho_{\scriptscriptstyle{\rm MF}} - \bar{\rho}^{(\varepsilon)}_{\scriptscriptstyle{\rm DMRG}} \bigr\|_F$, quantifies the discrepancy between the mean-field approximation and the \gls{dmrg} ground state.
This contribution depends on the quality of the mean-field description and scales as $\mathcal{O}(\sqrt{\Delta})$.
We now turn to the convexity properties of $D_F$. 
The cost function constitutes a convex quadratic optimization problem with respect to $\rho^{(\varepsilon)}_{\scriptscriptstyle{\rm MF}}$, as can be seen by rewriting it as
\begin{align}
    D_F
    &= \bigl\| \rho^{(\varepsilon)}_{\scriptscriptstyle{\rm MF}} - \rho^{(\varepsilon)}_{\scriptscriptstyle{\rm DMRG}} \bigr\|_F^2 \\
    &= \mathrm{tr}\!\left\{
        \left(\rho^{(\varepsilon)}_{\scriptscriptstyle{\rm MF}} - \rho^{(\varepsilon)}_{\scriptscriptstyle{\rm DMRG}}\right)
        \left(\rho^{(\varepsilon)}_{\scriptscriptstyle{\rm MF}} - \rho^{(\varepsilon)}_{\scriptscriptstyle{\rm DMRG}}\right)^{\dagger}
    \right\} \nonumber \\
    &= \mathrm{tr}\!\left\{ \left(\rho^{(\varepsilon)}_{\scriptscriptstyle{\rm MF}}\right)^2 \right\}
    + \mathrm{tr}\!\left\{ \left(\rho^{(\varepsilon)}_{\scriptscriptstyle{\rm DMRG}}\right)^2 \right\}
    - 2\,\mathrm{tr}\!\left\{ \rho^{(\varepsilon)}_{\scriptscriptstyle{\rm MF}} \rho^{(\varepsilon)}_{\scriptscriptstyle{\rm DMRG}} \right\},
\end{align}
where we have used the linearity and cyclicity of the trace.
The first term is a positive semidefinite quadratic form in the entries of $\rho^{(\varepsilon)}_{\scriptscriptstyle{\rm MF}}$, the second term is constant, and the final term depends linearly on $\rho^{(\varepsilon)}_{\scriptscriptstyle{\rm MF}}$.
Consequently, $D_F$ is convex in $\rho^{(\varepsilon)}_{\scriptscriptstyle{\rm MF}}$.
In practice, the density matrix $\rho^{(\varepsilon)}_{\scriptscriptstyle{\rm MF}}$ depends on the mean-field parameter $\Phi$ through the ground state of the impurity Hamiltonian $\hat{H}_{\rm imp}(\Phi)$. 
While this renders $D_F(\Phi)$ generally non-convex over the full parameter space, near the optimal solution $\rho^{(\varepsilon)}_{\scriptscriptstyle{\rm MF}}(\Phi)$ varies smoothly with $\Phi$ and can be linearized. 
As a result, $D_F(\Phi)$ is locally quadratic and convex in the vicinity of the minimum, thereby justifying the uniqueness and stability of the optimal mean-field solution reported in the main text.
\end{appendix}
\end{document}